\documentstyle[pre,aps,epsf]{revtex}
\newcommand{\cO}{{\cal{O}}}
\newcommand{\BEQ}{\begin{equation}}
\newcommand{\EEQ}{\end{equation}}
\newcommand{\BEA}{\begin{eqnarray}}
\newcommand{\EEA}{\end{eqnarray}}
\renewcommand{\H}{{\cal {H}}}

\renewcommand{\S}{S_{\em ep}}

\newcommand{\I}{{\cal {I}}}
\newcommand{\nn}{\nonumber }
\newcommand{\Kt}{{\tilde K}}
\newcommand{\Ht}{{\tilde H}}

\newcommand{\bam}{{\overline m}}

\newcommand{\bamu}{{\overline {\mu}}_2}
\newcommand{\tinf}{{\mbox{\hspace*{-0.2 mm}\tiny $\infty$}}}

\renewcommand{\thesection}{\arabic{section}}

\renewcommand{\theequation}{\thesection\arabic{equation}}

\begin{document}
\title{Inherent Structures in   models for fragile and strong   glass}
\author{Luca Leuzzi   and    Theo~M.~Nieuwenhuizen\\
Universiteit van Amsterdam
\\ Valckenierstraat 65, 1018 XE Amsterdam, The Netherlands}
\date{  printout: \today}
\maketitle
\begin{abstract}
An analysis of the dynamics is performed of
 exactly solvable models  for fragile and strong glasses,
exploiting the partitioning of the free energy landscape in inherent 
structures.
The results  are compared with the exact solution
of the dynamics,
by employing the formulation of an effective temperature used in literature. 
Also a new formulation is   introduced, based 
upon general statistical considerations, that 
 performs better.
Though the  considered models are conceptually simple there is no limit 
in which the inherent structure approach is exact.

\end{abstract}

\renewcommand{\thesection}{\arabic{section}}
\section{Introduction}
\setcounter{equation}{0}\setcounter{figure}{0}
\renewcommand{\thesection}{\arabic{section}.}
\label{intro}

The characteristics of a glassy system \cite{ANGELL,MCKENNA}
 arises from
the complex topography of the multidimensional function representing the 
collective potential energy that gives rise to a
non-trivial partition function and thermodynamic potential.
In this picture, at low enough temperature where vibrations are minimal,
 the spatial atomic patterns in crystals and in 
amorphous systems
share the common basic attribute that both represent minima in the potential
energy function describing the interactions.
The  presence of distinct processes acting on
two different time-scales means that the deep and wide local minima 
 at and below the glass transition temperature 
$T_g$ are geometrically organized to 
create a two scale length potential energy pattern.
$T_g$ depends on the cooling procedure and it is usually determined as the 
temperature at which the viscosity of the glass former
reaches the value of $10^{13}$ Poise.

In the present paper we  investigate,  using the inherent structure approach,
an exactly
 solvable  model glass that shows all the basic  features
of  real glasses \cite{LN}.
The model is  built 
 by processes evolving on two different, well separated time scales,
 representing respectively 
the $\alpha$ and $\beta$ processes taking place in real
glassy materials.
The slow $\alpha$ processes represents the
 escape from one deep minimum within a large scale 
valley to another valley. The fast $\beta$ processes, instead, are 
related  to elementary relaxations between neighbouring
minima inside the same valley.
We consider here all kinds of $\beta$ processes as equivalent, 
since the  characteristic time-scales  on which they are evolving
 are in any case much shorter than the time-scale of the $\alpha$
processes (i.e. the observation time).

In the general case, 
decreasing the temperature, the  free energy local minima 
can, in principle, be split into smaller
 local minima. 
But if we can  assume that they maintain their 
identity in spite of this splitting, we can set a one-to-one 
correspondence between local minima and inherent structures 
\cite{SW82,SW84,SW95,SADEB}, i. e. between the minima of the
free energy and the ones of the potential energy.
Actually, such a splitting is not even present in the two dynamical
models presented here after, making the correspondence clearer.

In this paper  we will 
 see to which extent such a scheme, widely used
in numerical simulations \cite{SW82,SW84,SW95,KST,SKT,CR1,CR2,CR3}, applies 
to our analytically solvable model. We will compare it  with the
exact dynamic solution, achieved without any partitioning of the
configuration space.

In section \ref{sec2} we do introduce the two kinetic models
and we give the description of their statics and of their
Monte Carlo dynamics.
In section \ref{sec3} we develop the inherent structure approach for 
the dynamics of such models  and we define two different 
inherent structure effective temperatures mapping the dynamics into a
 thermodynamic frame (in section \ref{sec3:Te}); 
one definition follows the  literature about numerical simulations
\cite{SW82,SW84,SW95,SADEB,KST,SKT,CR1,CR2,CR3}, 
the other exploits the 
analytic solubility of the models.



 \renewcommand{\thesection}{\arabic{section}}
 \section{ The models and their properties }
\setcounter{equation}{0}\setcounter{figure}{0}
\renewcommand{\thesection}{\arabic{section}.}
\label{sec2}

\subsection{Hamiltonian and its constraint}
We present two dynamical models, having the same statics,
but different dynamics bringing to the behavior of a fragile glass in one
instance and to the behavior of a strong glass in the other one.  
The version describing a system relaxing like a fragile glass was
introduced in \cite{NCM} and widely studied in
\cite{LN}.

Both models are
 described by the following local Hamiltonian:
\BEQ
{\cal{H}}[\{x_i\},\{S_i\}]=\frac{1}{2} K
\sum_{i=1}^{N}x_i^2
-H\sum_{i=1}^{N}x_i
-J\sum_{i=1}^{N}x_iS_i
-L\sum_{i=1}^{N}S_i
\label{Hmodel}
\EEQ

\noindent where $N$ is the size of the system and 
$\{x_i\}$ and $\{S_i\}$ are continuous variables, 
the last satisfying a spherical constraint: $\sum_i S_i^2=N$.
We  call them respectively harmonic oscillators
and spherical spins.
$K$ is the Hooke elastic constant, $H$ is an external field acting on
the harmonic oscillators, $J$ is the coupling constant between $\{x_i\}$ 
and $\{S_i\}$ on the same site $i$
 and $L$ is the external field acting on the spherical spins.
 A separation
of time scales
 is introduced by hand: the spins represent the fast modes and  the
harmonic oscillators the slow ones.
We  assume that the $\{S_i\}$ relax to equilibrium
 on a 
time scale  much shorter
than the one of the harmonic oscillators.
From the point of view of the motion of the $\{x_i\}$, the spins are just
a noise. 
To describe the long time regime of the $\{x_i\}$, in \cite{LN},
 we did average over this noise by performing the computation 
of the $\{S_i\}$ partition function,
obtaining an effective Hamiltonian depending only on the $\{x_i\}$,
that  determines the dynamics of these variables.
 Using the 
saddle point approximation for large $N$  we got:
\BEA
Z_S(\{x_i\})&=&
\int\left(\prod_{i=1}^{N} dS_i\right)
 \exp\left\{-\beta{\cal H}\left[\{x_i\},\{S_i\}\right]\right\}\,\,
\delta\left(\sum_{i=1}^{N}S_i^2-N\right) 
\label{Z_S}
\\ \nn
&\simeq&
\exp\left[-\beta N\left(
\frac{K}{2}  m_2  - H  m_1  - w +
 \frac{T}{2}\log\frac{w+\frac{T}{2}}{T} \right)\right]
\EEA
Where we introduced the short-hands
\BEQ
m_1\equiv\frac{1}{N}\sum_{i=1}^Nx_i \hspace*{2 cm}
m_2\equiv\frac{1}{N}\sum_{i=1}^Nx_i^2 \ 
\label{abbrev_m}
\EEQ
and 
\BEQ
w\equiv\sqrt{J^2 m_2+2JLm_1+ L^2+ \frac{T^2}{4}} \ .
\label{def:w}
\EEQ        
We  define, then, the effective Hamiltonian
${\cal{H}}_{\rm eff}(\{x_i\})\equiv -T\log Z_S(\{x_i\})$, that is the free
energy for a given configuration of $\{x_i\}$. We find
\BEQ               
{\cal{H}}_{\rm eff}(\{x_i\})=
\frac{K}{2}  m_2 N - H  m_1 N - w N+
 \frac{TN}{2}\log\frac{w+\frac{T}{2}}{T}
\label{Heff}
\EEQ                                                             
This can also be written  in terms of the internal energy $U(\{x_i\})$
and of the entropy $S_{\rm ep}(\{x_i\})$
of the equilibrium processes (i.e. the spins):
\BEA
{\cal{H}}_{\rm eff}(\{x_i\})&=&U(\{x_i\})-T S_{\rm ep}(\{x_i\})
\label{def:Heff}\\
U(\{x_i\})&=& \frac{K}{2}  m_2 N - H m_1 N - w N+ \frac{TN}{2}
\label{def:U}\\
S_{\rm ep}(\{x_i\})&=&\frac{N}{2}-\frac{N}{2}\log \frac{w+T/2}{T}
\label{def:Sep}
\EEA
\noindent The function $U$ is  actually the Hamiltonian 
averaged over the spins and $S_{\rm ep}$ is the entropy of the spins. 

In \cite{LN} we studied
the  model  characterized  by 
a constraint on the phase space, introduced for the fragile glass case 
 to avoid the existence of the single 
global minimum, and 
implementing a large degeneracy of the allowable lowest states. 
The constraint is taken on the $\{x_i\}$,
 thus concerning the long time regime. It reads:
\BEQ
m_2-m_1^2\geq m_0 
\label{CONSTRAINT}
\EEQ
where $m_0$ is a model parameter. 
It is a  fixed but arbitrary, strictly positive constant.
This constraint applied to the harmonic oscillators dynamics is
a way to reproduce the behavior of good glass formers.
We imposed 
a Monte Carlo dynamics \cite{BPPR,BPR}
satisfying this constraint and coupling 
the otherwise non-interacting $\{x_i\}$ in a dynamic way.
As we saw in \cite{LN},
the system exhibits a
Vogel-Fulcher-Tammann-Hesse (VFTH) relaxation
\cite{VFTH}, characterizing a fragile glass.

To model a strong glass, instead,
 we will also consider the same model Hamiltonian
but without  imposing any constraint
and  making  use of  a different
Monte Carlo dynamics.
We will show later in this paper that this dynamics displays an Arrhenius
relaxation  near zero temperature. In this case we 
have  a strong glass, 
as it happens for similar models, e. g. the oscillators model \cite{BPR}
and the spherical spins model \cite{NPRL98} where exactly the same 
dynamics is applied.
The new point of the present model is that now
both fast and slow processes occur.

To shorten the notation 
 we define  the modified ``spring constant'' 
${\tilde{K}}$ and ``external field'' ${\tilde{H}}$:
\BEQ
{\tilde{K}}=K-\frac{J^2}{w+T/2},\hspace*{2 cm}
{\tilde{H}}=H+\frac{JL}{w+T/2}
\label{def:tilde}
\EEQ
We stress  that ${\tilde{K}}$ and ${\tilde{H}}$ are actually functions of
the $\{x_i\}$ themselves (through $m_1$ and $m_2$ that occur  in $w$).
We also define the constant
\BEQ
D\equiv HJ+KL. 
\EEQ  
Using the definitions
 (\ref{def:tilde})
it is useful  to note 
that 
\BEQ
{\tilde{H}}J+{\tilde{K}}L=HJ+KL=D.
\label{Drel}
\EEQ


\subsection{Statics at heat-bath temperature $T$}

The partition function of the whole system at equilibrium is:
\BEA
Z(T)&&=\int {\cal{D}}x {\cal{D}} S \exp\left[-\beta{\cal{H}}(\{x_i\},\{S_i\})\right] \delta\left(\sum_i x_i^2-N\right)=
\nn
\\
\label{Z_static}
&&=
\int d m_1 d m_2 \exp\left\{-\beta N\left[\frac{K}{2} m_2 -H m_1-w+\frac{T}{2}\log\left(\frac{w+T/2}{T}\right)-\frac{T}{2}\left(1+\log(m_2-m_1^2)\right)\right]\right\}
\EEA
The new object that appears in the exponent is the 
configurational entropy
\BEQ
{\cal{I}}\equiv\frac{N}{2}\left(1+\log(m_2-m_1^2)\right)
\label{def:I}
\EEQ
It comes  from the 
Jacobian $\exp{\I}$ of the transformation of variables
${\cal{D}}x \to dm_1dm_2$,
(see (\ref{abbrev_m})).
We can compute the large $N$-limit of this partition function using once again
the saddle point 
approximation.
The saddle point equations are found minimizing 
the expression between square brackets in (\ref{Z_static})
with respect to $m_1$ and $m_2$.
This yields
\BEA
\bam_1&&=\frac{\Ht(\bam_1,\bam_2)}
{\Kt(\bam_1,\bam_2)}
\label{SPeq1}
\\
\bam_2&&=
\bam_1^2+
\frac{T}
{\Kt(\bam_1,\bam_2)}\label{SPeq2}
\EEA
The form of the solutions $\bam_1(T)$, $\bam_2(T)$
 is quite complicated because each of these
equations is actually a fourth order equation, but they can 
be explicitly computed.
In terms of the equilibrium values $\bam_k$ we find the following expression for the equilibrium free energy:
\BEA
F(T,\bam_1(T),\bam_2(T))&&=N\left\{\frac{K}{2} \bam_2 -H \bam_1-
w\left(\bam_1,\bam_2\right)
+\frac{T}{2}\left[\log
\frac{w\left(\bam_1,\bam_2\right)+T/2}{T}
-\left(1+\log(\bam_2-\bam_1^2)\right)\right]\right\}
\\
&&=\hspace*{ 1 cm} U(T,\bam_1,\bam_2)\hspace*{ 1.15 cm}  -
\hspace*{ 1.15 cm}   T \ S_{\rm ep}(T,\bam_1,\bam_2)
\hspace*{ 0.5 cm}  -
\hspace*{ 0.5 cm} T \ {\cal{I}}(T,\bam_1,\bam_2)
\EEA

This is the statics both for the model with the constraint 
(\ref{CONSTRAINT}), as long as the temperature exceeds 
the Kauzmann temperature, and
for the one without it.
Indeed for  the fragile glass case
at $T\leq T_0$, when the constraint is reached, the saddle point
equation (\ref{SPeq2}) becomes $\bam_2-\bam_1^2=m_0$, no matter what the 
temperature of the thermal bath is.
In this work, however, we will limit ourselves for the 
fragile glass to cases where $T$ is slightly larger then $T_0$,
and for the strong glass case to temperature slightly above zero.


\subsection{Dynamics}
\label{dyn}
The dynamics we apply to the system is a
parallel Monte Carlo dynamics firstly introduced
 in \cite{BPPR}.
 The thus obtained dynamical
model composed by the  simple local  Hamiltonian (\ref{Hmodel})
and such a dynamics has the
benefit of being analytically
solvable.

In a Monte Carlo step a random updating of the variables is performed
($x_i\to x'_i=x_i+r_i/\sqrt{N}$) where the $\{r_i\}$ have a Gaussian 
distribution with zero mean and variance $\Delta^2$.
We define $x\equiv{\cal{H}}(\{x'_i\})-{\cal{H}}(\{x_i\})$ as
the energy difference 
between the new and the old state.
If $x>0$ the move is 
 accepted with a probability $W(\beta x)\equiv
\exp(-\beta x)$; else 
 it is  always accepted ($W(\beta x)=1$).
The updating is made in parallel.
It is the parallel nature of the updating that allows
 the collective behavior leading to exponentially divergent time scales
in  models with no  interactions between particles such us ours.
A sequential updating would not produce any glassy effect.
 This dynamics may induce glassy behavior in situations 
where ordinary Glauber dynamics \cite{GLAUBER} would not. 
In our model the parallel  dynamics mimics the presence of interactions
between atoms in realistic glasses, where a high internal cooperativeness is present.
For different examples of dynamics implying non trivial collective behavior
the reader can look, for instance, at the  $n$-spin facilitated kinetic 
Ising model \cite{FA,FR} or at the kinetic lattice-gas model \cite{KA,KPS}.

In a Monte Carlo step the quantities $Nm_1=\sum_ix_i$ and 
$Nm_2=\sum_ix_i^2$ are updated. We denote their change by 
$y_1$ and $y_2$, respectively. Their  distribution function is, 
for given values   of $m_1$ and $m_2$,
\BEA
p(y_1,y_2|m_1,m_2)&\equiv &\int\prod_i \frac{d r_i}{\sqrt{2\pi\Delta^2}}
e^{-{r_i^2}/({2\Delta^2})}\,\,
\delta\left(\sum_i {x'}_i-\sum_ix_i-y_1\right)\,
\delta\left(\sum_i {x'}_i^{2}-\sum_ix_i^2-y_2\right)
\nn \\ &=&
\frac{1}{4\pi \Delta^2\sqrt{m_2-m_1^2}}
\exp\left(-\frac{y_1^2}{2\Delta^2}-\frac{(y_2-\Delta^2-2y_1m_1)^2}
{8\Delta^2(m_2-m_1^2)}\right)
\label{Ptrans}
\EEA
Neglecting the variations of $m_1$ and $m_2$ 
 of order $\Delta^2/N$ we can express the energy difference as \cite{LN}
\BEQ
x=\frac{\Kt}{2} \ y_2 -{\tilde{H}} \ y_1,
\label{x}
\EEQ

In terms of $x$
and $y=y_1$ the distribution function can be formally written
 as the product of two other Gaussian  distributions:
\BEA
p(y_1,y_2|m_1,m_2)dy_1 dy_2
	&=&\hspace*{ 10 mm}dx\hspace*{5 mm}p(x|m_1,m_2)
\hspace*{ 1.5 cm} dy\hspace*{5 mm}p(y|x,m_1,m_2)
\nn\\
&=&\frac{dx}{\sqrt{2\pi \Delta_x}}
	\exp\left(-\frac{(x-{\overline{x}})^2}{2 \Delta_x}\right)
	\frac{ dy}{\sqrt{2\pi \Delta_y}}
	\exp\left(-\frac{(y-{\overline{y}}(x))^2}{2 \Delta_y}\right)
	\label{PROBDIST}
\EEA

\noindent where 
\BEA
{\overline{x}}=\Delta^2{\tilde{K}}/2, \hspace*{2.5 cm} &&  \hspace*{2 cm}
\Delta_x=\Delta^2{\tilde{K}}^2
(m_2-m_1^2)+\Delta^2{\tilde{K}}^2\left(m_1-{\tilde{H}}/{\tilde{K}}\right)^2,\\
 {\overline{y}}(x)=\frac{m_1-{\tilde{H}}/{\tilde{K}}}{
m_2-m_1^2+\left(m_1-{\tilde{H}}/{\tilde{K}}\right)^2}
\frac{x-{\overline{x}}}{\tilde{K}},
 &&  \hspace*{2 cm}
\Delta_y=\frac{\Delta^2(m_2-m_1^2)}
{m_2-m_1^2+\left(m_1-{\tilde{H}}/{\tilde{K}}\right)^2}.
\label{def:MCave}
\EEA

\subsubsection{Dynamics of the fragile glass model}
To represent a fragile glass the dynamics that 
we apply to the system is a generalization of the
analytic treatment of
 Monte Carlo dynamics introduced
 in \cite{BPPR}.
As noted in \cite{NCM}, also in this generalized case  the  dynamical
model with  a contrived dynamics can be
 analytically solved. As we saw in \cite{LN},
in the long-time domain, the dynamics looks
quite reasonable with regard to what one might expect of any glassy
system and the system exhibits a
VFTH relaxation.
We repeat here the main steps of the implementation of this dynamics
(for a more extended presentation see \cite{LN}).

We let $\Delta^2$, the variance 
 of the random updating $\{r_i\}$,
  depend on the distance from the 
constraint, i.e. on  the whole $\{x_i\}$ configuration before the 
Monte Carlo update:

\BEQ
\Delta^2(t)\equiv 8[m_2(t)-m_1^2(t)]
\left(\frac{B}{m_2(t)-m_1^2(t)-m_0}\right)^\gamma
\label{Delta}
\EEQ
where $B$, $m_0$ and $\gamma$ are constants.
In particular  $\gamma$ is an exponent larger than zero
that appears in the VFTH-like relaxation
law of the model, when $T$ decreases towards some critical temperature $T_0$
(identified in \cite{LN} with the Kauzmann temperature):
\BEQ
\tau_{\rm eq}\sim \exp\left(\frac{A_{\rm f}}{T-T_0}\right)^\gamma
\label{tauF}
\EEQ
\noindent where $A_{\rm f}$ is a  constant depending on the system's parameters.
In other models \cite{BPPR,BPR,NPRL98,N00}
the variance $\Delta^2$  was  kept constant. The same  will also be  done 
for the strong glass version of the model in the next subsection.

For what concerns the exponent $\gamma$ we saw in \cite{LN}
that it generates different dynamic regimes for $\gamma>1$, 
$\gamma=1$ and $0<\gamma<1$;
  the situation $\gamma=1$ remains
 model dependent even in the long time limit.
We will stay in the following in the regime for $\gamma>1$.

The nearer the system goes to the constraint (i.e. the smaller the 
value of $m_2-m_1^2-m_0$), the larger the variance $\Delta^2$ becomes, 
implying almost always a refusal of the proposed updating.
In this way, in the neighborhood of the constraint,  the dynamics 
is very slow  and goes on through  very seldom but very
large moves, that can be interpreted as activated processes.
When the constraint is reached the variance 
$\Delta^2$ becomes infinite and the system
 dynamics gets stuck.
The system does not evolve anymore towards equilibrium
 but it is blocked in one  single ergodic component
of the configuration space. At large enough temperatures,
the combination $m_2(t)-m_1^2(t)-m_0$ will remain strictly positive.
The highest temperature, $T_0$, at which it can vanish for
$t\to\infty$, is identified with the Kauzmann temperature \cite{LN}.

In \cite{LN} the  dynamics was expressed in terms of   two combinations of
 $m_1$ and $m_2$. 
The first one, defined  as 
\BEQ
\mu_1\equiv\frac{\tilde{H}}{\tilde{K}}-m_1.
\label{defMU1}
\EEQ
\noindent  represents the distance from
the instantaneous equilibrium 
state.
 By instantaneous equilibrium state we mean that $\Ht$ and $\Kt$
 depend on the  values of $m_1$ and $m_2$ at a given time $t$.
For $t\to \infty$, at the true equilibrium, one has $\mu_1=0$.

The second dynamical variable  is defined as the distance from 
the constraint (\ref{CONSTRAINT}):
\BEQ
\mu_2\equiv m_2-m_1^2-m_0.
\EEQ
When $\mu_2=0$ the constraint is reached. This will happen if the temperature
is low enough ($T\leq T_0$) and the time large enough. 
$T_0$ is the highest temperature at which
the constraint is asymptotically ($t\to\infty$) reached by the system.
Above $T_0$ ordinary equilibrium will be achieved without reaching the constraint. 
The temperature is, then,  too high for the system to notice that there is a 
constraint at all on the  configurations, and this implies (see (\ref{SPeq2})):
\BEQ 
\lim_{t\to \infty}\mu_2(t)= \bamu(T)=\frac{T}{\Kt_{\tinf}(T)}-m_0>0 \ ,
\label{bamu2}
\EEQ
\noindent where 
\BEQ
 \lim_{t \to\infty}\Kt\left(m_1(t,m_2(t);T\right)\equiv\Kt_{\tinf}(T)=
\Kt\left(\bam_1(T),\bam_2(T)\right)  .
\EEQ
Below $T_0$ the system goes to configurations that become arbitrarily
close to the constraint, and then stay there arbitrarily long.
Note that, by definition of $T_0$ we can write
\BEQ
m_0=\frac{T_0}{\Kt_{\tinf}(T_0)}
\label{m0T0}
\EEQ

Solving the equations of motions, for fixed parameters
(aging setup), 
we find, to the leading orders of approximation for large times,
the following behavior for $\mu_2$ \cite{LN}:
\BEQ
\mu_2(t)\simeq \frac{B}{\left[\log (t/t_0)+
c\log\left(\log(t/t_0)\right)\right]^{1/\gamma}} 
\label{MU2F}
\EEQ 
\noindent where $c=1/2$  since in this paper we  only look at
the regime for  $T\geq T_0$.
The constant $t_0$ depends on the parameters of the model
 and on the temperature;
it is  of order one.
The solution (\ref{MU2F}) is valid in the aging regime, where 
$t_0\ll t\ll\tau_{eq}(T)$.
Indeed, when $t\sim\tau_{eq}(T)\sim \exp\left(A/(T-T_0)\right)^{\gamma}$
 the ``distance'' $\mu_2$ becomes
\BEQ
\mu_2\simeq\frac{B}
{\left[\left(\frac{A_{\rm f}}{T-T_0}\right)^{1/\gamma}\right]^{\gamma}}
\propto T-T_0 \ ,
\EEQ
\noindent as it should be.

We also introduce another variable that will be useful later on, namely
 the difference
between $\mu_2(t)$ and its asymptotic, equilibrium, value $\bamu(T)$:
\BEQ
\delta\mu_2(t)\equiv \mu_2(t)-\bamu(T)\simeq
\frac{B}{\left[\log (t/t_0)+
c\log\left(\log(t/t_0)\right)\right]^{1/\gamma}} 
-\frac{T}{\Kt_{\tinf}(T)}+\frac{T_0}{\Kt_{\tinf}(T_0)}
\label{deltamu2F}
\EEQ
\noindent where, using   (\ref{m0T0}), $\bamu(T)$ comes from (\ref{bamu2}), 
valid, in the fragile case, when $T\geq T_0$.
When $t\to \infty$ is, by definition, $\delta\mu_2=0$.

The dynamical behavior of $\mu_1$ depends not only on the temperature 
(above or below $T_0$) but also  on $\gamma$ being greater, equal to or lesser
 than one.
With respect to the relative weight of $\mu_1$ and 
$\mu_2$ we can identify different regimes, where the solution has
different behaviors \cite{LN}. What is of our interest here is the regime 
of $T\geq T_0$ and  $\gamma>1$,
 where $\mu_1(t)\ll\mu_2(t)$  and
a unique effective thermodynamic parameter
 can be properly defined in various
independent ways \cite{LN}.

\subsubsection{Dynamics of the strong glass model}
\label{dynS}
We now analyze the simple case without constraint on the configuration space
 and where $\Delta^2$, 
the variance 
 of the randomly chosen updating $\{r_i\}$
of the slow variables $\{x_i\}$, is a constant. This dynamical 
model can also be seen 
as the limit for $m_0\to 0$ and $\gamma\to 1$ of the preceding one.
We also mention that the case with $J=L=0$ is the model 
of harmonic oscillators studied in \cite{BPR,N00}.

In the fragile glass case
 we studied  a different version of such a 
  dynamics for two particular combinations of
the variables $m_1$ and $m_2$. Here we will keep the same notation.
The first variable 
is thus defined, starting from the saddle point equation (\ref{SPeq1}),
as the deviation from the instantaneous equilibrium 
state and is formally equivalent to 
(\ref{defMU1}).

The second  variable is defined as
\BEQ
\mu_2\equiv m_2-m_1^2.
\label{defMU2}
\EEQ
When $T=0$ from equation (\ref{SPeq2}) we know that
$\mu_2=0$.
Indeed at $T=0$ the system reaches its minimum
\BEQ
x_i=\frac{H+J}{K} \hspace*{1 cm}\forall i\ .
\label{minS}
\EEQ
\noindent For simplicity we limit ourselves to a choice of
the interaction parameters such that  $D=H J+K L>0$ and $\Kt>0$, for which this is the 
global minimum.
In appendix A we derive the equations of motion for 
$\mu_1$ and $\mu_2$ and we solve them for temperature equal to
and slightly above  zero
and long  times, in the aging regime.
In this time regime  $\mu_1$ comes out to be
much smaller than $\mu_2$: $\mu_1\propto\mu_2^2$. 
The solution for $\mu_2$ comes out to be, at the leading order
\BEQ
\mu_2(t)\simeq
\frac{\Delta^2}{8}
\frac{1}
{\log \frac{2 t}{\sqrt{\pi}}}
\label{MU2S}
\EEQ
The difference between $\mu_2(t)$ and its asymptotic value is now:
\BEQ
\delta\mu_2(t)\equiv \mu_2(t)-\bamu(T)\simeq
\frac{\Delta^2}{8}
\frac{1}
{\log \frac{2 t}{\sqrt{\pi}}}
-\frac{T}{\Kt_{\tinf}(T)}
\label{deltamu2S}
\EEQ
\noindent where $\bamu(T)$ comes from (\ref{SPeq2}) and 
\BEQ
  \Kt_{\tinf}(T)=\lim_{t\to\infty}\Kt\left(m_1(t),m_2(t);T\right)=
\frac{K D}{D+J^2}
+\frac{T}{2}\frac{J^2K^2}{(D+J^2)^2}
+\frac{T^2}{8}\frac{J^6K^3(J^2-3 D)}{D(D+J^2)^5}
+\cO(T^3)
\label{KtSexp}
\EEQ
For $t\to \infty$, $\delta\mu_2(t)\to 0$.

At low temperature,
the relaxation time for the slow processes depends on the temperature 
following an Arrhenius law:
\BEA
&&\tau_{\rm eq}(T)\propto \exp\left(\frac{A_{\rm s}}{T}\right) \ ;
\label{tauS}
\\
&&A_{\rm s}\equiv \frac{\Delta^2 \Kt_{\tinf}(0)}{8}\ .
\EEA

\subsection{Two temperature thermodynamics}
\label{sec2:2T}
Before going on we recall here that we are able to introduce
  effective parameters
in order to rephrase the dynamics of the system out of equilibrium
into a thermodynamic description (for a review see \cite{N00}). 

In \cite{LN} we got through different methods the following expression
for the effective temperature in the regime for $T>T_0$ as a function of the 
interaction parameters of the model and of the time evolution
of its observables:
\BEQ
T_e(t)=\Kt\left(m_1(t),m_2(t)\right)\left[m_0+\mu_2(t)\right] \ .
\label{Te_T}
\EEQ

Since we will use  one of these methods  in the next section
to map the IS dynamics into an effective thermodynamic parameter
 we shortly recall this particular  derivation of (\ref{Te_T}).
 Knowing the solution of the dynamics at a given time $t$
a quasi-static approach 
can be followed by  computing the 
 partition function $Z_e$ of all the macroscopically 
equivalent states 
at the time  $t$.
In order to generalize the equilibrium thermodynamics we 
assume an effective temperature $T_e$ and an effective field $H_e$,
and substitute the Boltzmann-Gibbs equilibrium measure by 
$\exp (-\H_{\rm eff}(\{x_i\},T,H_e)/T_e)$, 
where $\H_{\rm eff}$ is given in (\ref{def:Heff})
 and the true external field $H$ in it has been substituted by
the effective field $H_e$.
As  we get the expression of the ``thermodynamic''
potential $F_e\equiv-T_e \log Z_e$  as a function
of macroscopic variables $m_{1,2}$ and effective parameters,
we can determine $T_e$ and $H_e$ 
minimizing $F_e$ with
respect to $m_1$ and $m_2$ 
and evaluating
the resulting analytic expressions at $m_{1,2}=m_{1,2}(t)$.

The partition function of the macroscopically equivalent states is:
\BEQ
Z_e\left(m_1,m_2;T_e,H_e\right)\equiv \int {\cal{D}}x 
\ \exp\left[-\frac{1}{T_e}\H_{\rm eff}(\{x_i\},T,H_e)
\right] \ 
\delta( N m_1-\sum_ix_i)\ \delta( N m_2-\sum_ix_i^2),\label{Ze}
\EEQ

From this  we  build the effective thermodynamic potential
as a function of $T_e$ and $H_e$, besides of $T$ and $H$, where
the effective parameters depend on time through the
time dependent values of $m_1$ and $m_2$, solutions of the dynamics. 
They are actually
a way of describing the evolution in time of the system out of equilibrium.
The  free energy $F_e=-T_e\log Z_e$ is minimized with respect
to  $m_1$ and $m_2$.
Then their time dependent  values are inserted, yielding
\BEQ
F_e(t)=U\left(m_1(t),m_2(t)\right)
-T\S\left(m_1(t),m_2(t)\right)-T_e(t)\I\left(m_1(t),m_2(t)\right)
+[H-H_e(t)]Nm_1(t), \label{Fe}
\EEQ
\noindent with 
\BEA
T_e(t)&=&\Kt\left(m_1(t),m_2(t)\right) \left[m_0+\mu_2(t)\right],
 \nn
\\
 H_e(t)& =&H-\Kt\left(m_1(t),m_2(t)\right) \mu_1(t) \ .
\label{He_T}
\EEA
\noindent  where $T_e$ is the so called effective temperature.
$U$ is the internal energy of the whole system,
$\S$ is the entropy  of the fast or equilibrium 
processes (the spherical spins) while ${\cal{I}}$ is the 
entropy of the slow, ''configurational'', processes
(the harmonic oscillators). 
 The last term of $F_e$ replaces the $-HNm_1$ occurring
in $U$ (see eq. (\ref{def:U})) by $-H_eNm_1$.
$U$, $\S$ and $\I$ 
are 'state' functions, in the sense that  they depend on the state
described by $T$, $T_e$, $H$ and, if needed,  $H_e$.
 In the framework where  only one  relevant  effective parameter $T_e$ stays,
 these functions do not depend on the path along which its
value has been reached.

As we saw in \cite{LN} for the VFTH relaxing model at $T>T_0$
and in appendix A for the Arrhenius relaxing case (\ref{mu1Ssol0}),
(\ref{mu1SsolT}),
the effective temperature alone is
enough for a complete thermodynamic description of the dominant
physical phenomena ($H_e=H$). The introduction of $H_e$ becomes
important only for second order corrections in $\delta\mu_2$.

 \renewcommand{\thesection}{\arabic{section}}
 \section{Inherent Structure approach}
\setcounter{equation}{0}\setcounter{figure}{0}
\renewcommand{\thesection}{\arabic{section}.}
\label{sec3}

The characteristics of a glassy system
can be represented by means of a  multidimensional  potential energy 
function with a complex topography.
The spatial patterns of atoms in crystals and in amorphous systems,
at low temperature,
  represent, then,  minima in the potential
energy function describing the interactions \cite{SW82,SW84}.

In the case of the  model (\ref{Hmodel}) all the complex
chemical properties of  real glass formers are not present,
but nevertheless the system exhibits several aspects of
 their  complex features (described in \cite{LN}),
indicating that our simple model is complicated enough 
for what concerns the description and the comprehension 
of the basic long time properties of a glass.

In a real glass  the 
presence of distinct processes (acting on
different time-scales) can be obtained from a careful analysis of 
the relaxation response  function above $T_g$.
We limit ourselves to a two time-scale approach.
This means that the deep and wide local minima 
 at and below $T_g$ are geometrically organized to 
create a two scale length potential energy pattern.
As a consequence the system shows 
$\alpha$ and $\beta$ processes.
The $\alpha$ processes represents the
 escape from one deep minimum within a large scale 
valley to another valley. This escape requires a lengthy directed sequence
of elementary transitions producing a 
very large activation energy.
Moreover, the high lying minima between any two valleys, among which the
system is making a transition, are very degenerate. 
This implies a large activation 
entropy for the {\em inter-basin} transition.
$\beta$ processes are instead 
related  to elementary relaxations between neighboring
minima ({\em intra-basin} dynamics).

We stress that in our models we put together all kind of $\beta$ processes
in our short  time-scale, since they are in
any case much shorter than the observation time considered.


\subsection{Decomposition of the partition function:
 introduction of inherent structures}
In this point of view an approximate approach to the problem is to 
divide the complicated multidimensional landscape of the (potential) 
energy in structures formed by
large deep basins and to describe the dynamics of the processes
taking places as intra-basin and inter-basin
\cite{SW82,SW84}.

More precisely one can define an I{\em{nherent}} S{\em{tructure}} (IS) as that
basin behind an actual configuration of the system evolving in time at 
some temperature  $T$ that is the minimum of the potential energy reached 
in an instantaneous quenching by the method of {\em{steepest descent}}.

The introduction of IS's allows, at low enough temperature ($T < T_g$),
a decomposition of the partition function into an IS part, connected to the
zero temperature 
 landscape corresponding to the configurations of the system at temperature T,
and a part connected to the thermal excitation of the configurations in
 a single minimum.

The probability that an equilibrium configuration at $T=1/\beta$
belongs to a basin associated with an IS structure with an  energy density
in the interval $[e, e+de]$ is \cite{SW82,SKT,CR1,CR2}
\BEQ
{\cal{P}}(e,T) de\ \propto \ \exp\left(-\beta N\left[ e-T s_{\tiny{c}}(e)+
f_v(e,T)\right]\right)  de
\label{PIS}
\EEQ
\noindent
where $s_{\tiny{c}}(e)$ is proportional to
 the logarithm of the number of IS's
existing at the energy level $e$ and $f_v(e,T)$ is the free energy 
of the configurations inside an 
IS at energy $e$ (related to a temperature $T$ system).
To derive the distribution (\ref{PIS}) in this form
the approximation is made that $f_v$ is computed as the average 
over all the IS's of energy $e$. 
This means that, by assumption, the shape of a basin depends
only on its energy level and on the temperature.
Enough below $T_g$ the further approximation 
can be made, that $f_v(e,T)\sim f_v(T)$, because fluctuations inside one
 IS are small \cite{KST,SKT}. 
The shape of the basin depends then only on the
temperature. 
All the internal (vibrational) states  of any IS
have the same (vibrational) free energy $f_v$ at given $T$.
 We anticipate, however, that in the present study we will not
carry out such an approximation for our models.

IS dynamics is significant, i.e. significantly represents the actual dynamics 
of the system at finite T, provided that  there is a 
one to one correspondence between IS's and real minima at finite temperature
and provided that  these IS's  are visited with
the same frequency with which the corresponding finite $T$ 
minima are visited.


\subsection{Inherent Structure approach in the Harmonic
 Oscillator-Spherical Spin
 model}
As we will see,
the model (\ref{Hmodel}) is built in such a way that every $\{x_i\}$ 
configuration  
is an inherent structure. Indeed, at a given $\{x_i\}$ configuration at
 finite $T$,
the $\{S_i\}$ are fast variables and they contribute to the
energy and to the other observables as a noise depending on temperature.
If we take away this contribution we do not actually change 
the  configurations of the minima of the slow variables.
In the case of the system without constraint on  the configuration
space, nor contrived dynamics, any $\{x_i\}$ configuration
is an inherent structure.
For what concerns the constrained model, instead,
certain configurations
are not allowed. Moreover the presence of the constraint (\ref{CONSTRAINT})
produces  (entropic)
 barriers higher than in the other case
 to get from a certain IS to a different one. That just means that the 
dynamics through the inherent structures is even slower in the fragile glass case that in the strong glass case.

First of all we have to define  the steepest descent procedure for the model.
We start 
 performing the minimization of 
\BEQ 
\H+\lambda\sum_{i=1}^{N}S_i^2-\lambda N \ ,
\label{perMIN}
\EEQ
\noindent where $\H$ is the Hamiltonian (\ref{Hmodel}) of the model
and where we implemented the spherical constraint $\sum_i S_i^2=N$
by using the Lagrange multiplier $\lambda$.

To get rid of the contribution of the spins, i.e. to get rid
of the fast modes, we minimize  (\ref{perMIN}) with respect to the $\{S_i\}$.
We get 
\BEQ
 S_i^{(\rm{min})}=\frac{\beta(Jx_i+L)}{2\lambda} \hspace*{1 cm}\forall i
\EEQ
Inserting this values for $S_i$ and  solving for $\lambda$ we, then, find
\BEQ
\lambda=\frac{\beta}{2} w_{\rm is}
\label{l_SP}
\EEQ
\noindent where $m_1$ and $m_2$ are defined in (\ref{abbrev_m}) and 
\BEQ
w_{\rm is}\equiv \sqrt{J^2m_2 +2JL m_1 +L^2} \ .
\label{def:wis}
\EEQ
Using (\ref{l_SP}) the 
minimum $\{S_i\}$ configuration for a given set of $\{x_i\}$ is, 
thus, given
by
\BEQ
 S_i^{(\rm min)}=\frac{J x_i+ L}{w_{\rm is}}\hspace*{1	 cm} \forall i \ .
\EEQ

Finally the expression (\ref{perMIN}) becomes
\BEQ
\H_{\rm is}\equiv N\left[\frac{K}{2}m_2-Hm_1-w_{\rm is}\right]
\EEQ
\noindent that is the energy function of the inherent structures.
Consequently the partition  sum over inherent structures 
is defined by
\BEQ
Z_{\rm is}=\int{\cal{D}}x \exp{\left[-\beta \H_{\rm is}\right]}=
\int dm_1 dm_2 
 \exp{\left(\I-\beta \H_{\rm is}\right)}
\label{ZIS}
\EEQ
\noindent Due to the minimization any explicit dependence on $T$ disappears
with respect to the effective Hamiltonian (\ref{Heff}).
In (\ref{Z_S}), integrating over the spins, instead
of minimizing with respect to them,
we also had an entropic term for the fast processes
 ($S_{\rm ep}$ given in (\ref{def:Sep}))
 and a slightly different internal energy
($N w$ instead of $N w_{\rm is}$).
Carrying out  steepest descent the entropic term vanishes
 (only the minimal configuration is taken into account)
and the inherent structure energy has no explicit dependence
on the temperature.
The configurational entropy for IS's comes from the 
Jacobian of the transformation of variables
${\cal{D}}x = e^{\I} dm_1\ dm_2$, 
(see (\ref{abbrev_m}) and
(\ref{def:I})).
It is the same of the 
finite $T$ case, since any allowed configuration $\{x_i\}$ is also an IS.

The static average of ${\cal{H}}_{is}$ is given by
\BEQ
E_{\rm eq}^{\rm is}(T)=
{\cal{H}}_{is}(\bam_{1,2}^{(\rm is)}(T))
\EEQ
where $\bam_{1,2}^{(\rm is)}(T)$ are the solutions of the
saddle point equations that we get in the IS case to compute (\ref{ZIS}),
in the limit of large $N$. The equations are
\BEA
\bam_1^{(\rm is)}&&=\frac{D}{J\left(K-J^2/w_{\rm is}\right)}-\frac{L}{J}=
\frac{\Ht_{\rm is}}{\Kt_{\rm is}}
\label{eq:m1spIS}\\
\bam_2^{(\rm is)}&&-(\bam_1^{(\rm is)})^2=\frac{T}{D}(J\bam_1^{(\rm is)}+L)
=\frac{T}{\Kt_{is}}
\label{eq:m2spIS}
\EEA
where we define
\BEQ
\Ht_{\rm is}\equiv H+\frac{JL}{w_{\rm is}} \hspace*{1.5 cm} ; 
\hspace*{1.5 cm} \Kt_{\rm is}\equiv K-\frac{J^2}{w_{\rm is}}
\EEQ
\noindent with $w_{\rm is}$ from (\ref{def:wis}).
The combination $\Ht_{\rm is} J+ \Kt_{\rm is} L=H J+ K L=D$ is, again, simple, as 
in (\ref{Drel}).

In the case at finite $T$ the static partition function (\ref{Z_static})
was
\BEQ
Z=\int dm_1dm_2\exp{\left({\cal{I}}-\beta{\cal{H}}_{\rm eff}\right)},
\EEQ
\noindent with ${\cal{H}}_{\rm eff}$ defined in (\ref{Heff}) and 
${\cal{I}}$ in (\ref{def:I}). 
The two saddle point equations are different from (\ref{SPeq1})
and (\ref{SPeq2}) valid in the realistic case, giving thus different
results: $\bam_{1,2}^{(\rm is)} \neq \bam_{1,2}$.
We note explicitly that $\bam_{1,2}^{(\rm is)}$ 
depends on $T$ even in the IS case.

Comparing the expressions so far obtained with those appearing  in the exponent
of the probability distribution (\ref{PIS}) we identify
the configurational entropy $N s_c$ with $\I$, as defined in (\ref{def:I}),
and the rest with 
\BEQ
N(e+f_v)=\H_{\rm eff}\left(\{x_i\}\right)=\H_{\rm is}\left(\{x_i\}\right)
+F_v\left(\{x_i\}\right)
\EEQ
\noindent where, as already told, $\H_{\rm is}\left(\{x_i\}\right)$
is  the IS internal energy  and 
from the difference $\H_{\rm eff}\left(\{x_i\}\right)
-\H_{\rm is}\left(\{x_i\}\right)=F_v(\{x_i)\}$ 
the   thermal free energy  of one IS turns out to be:
\BEQ
F_{v}=\frac{T}{2}\log\left(\frac{w+T/2}{T}\right)- N(w-w_{is})\ .
\EEQ
\noindent where $w$ is defined in (\ref{def:w})
 and $w_{\rm is}$ in (\ref{def:wis}).
Notice that it explicitly  depends on the parameters $m_1$ and $m_2$ 
of the IS, whereas in literature it is often assumed to be a constant
(harmonic approximation \cite{SW82,SW84,KST,SKT,CR1,CR2}.


\subsection{Effective temperature in the IS's approach}
\label{sec3:Te}
\subsubsection{Expansion of the dynamical energy}

A possible way of defining an effective temperature, 
sometimes used in literature, for instance  in the study 
of Lennard-Jones interacting spheres \cite{KST,SKT} and in the study
 of the 
random orthogonal model \cite{CR1}, is to compare the time 
dependent
 out of equilibrium mean internal energy  with the
equilibrium mean internal energy expression at a temperature $T_e \neq T$.
The  out of equilibrium  mean internal energy
is built   taking   the dynamics of a system out of equilibrium 
at temperature $T$ and repeating it many times 
starting from different initial conditions.
A statistical ensemble of trajectories is constructed in this way. 
At any given 
time $t$  the configurations that each sample is visiting are found.
The energy ${\cal{H}}_{\rm is}$ averaged over the ensemble 
of different trajectories is,
\BEA
E_{d}^{(\rm is)}(t)&&\equiv\left<{\cal{H}}_{is}\right>_t
\equiv N\frac{K}{2}m_2(t)-NHm_1(t)-N\sqrt{J^2m_2(t)+2JLm_1(t)+L^2}
\label{Ed}
\\
&&
\simeq N\frac{K}{2}\bam_2^{(\rm is)}-NH\bam_1^{(\rm is)}
-N\sqrt{J^2\bam_2^{(\rm is)}+2JL\bam_1^{(\rm is)}+L^2}
+N\Kt_{is}(\bam_1^{(\rm is)},\bam_2^{(\rm is)})\delta\mu_2(t)+
C(\bam_1^{(\rm is)},\bam_2^{(\rm is)}) \delta\mu_2(t)^2\\
&&
\simeq\hspace*{1 cm} E_{\rm eq}^{(\rm is)}(T) \hspace*{1.5 cm} +
\hspace*{1.5 cm}N\Kt_{\rm is}(\bam_1^{(\rm is)},\bam_2^{(\rm is)})\delta\mu_2(t)
 \hspace*{1.5 cm} +
\hspace*{1.5 cm} C(\bam_1^{(\rm is)},\bam_2^{(\rm is)})\delta\mu_2(t)^2
\EEA
\noindent where $\delta \mu_2(t)\equiv \mu_2(t)-\bamu(T)$ is given by 
(\ref{deltamu2S}) in the Arrhenius case and by (\ref{deltamu2F})
 in the constrained  case for $T\geq T_0$.
The equilibrium IS energy $E_{\rm eq}^{(\rm is)}(T)$ will be a different 
function of the temperature in the two dynamic versions of 
the model. The second order expansion will be needed only for the strong glass case
and the expression for the 
factor $C(\bam_1^{(\rm is)},\bam_2^{(\rm is)})$ is, in that case:
\BEQ
C(\bam_1^{(\rm is)},\bam_2^{(\rm is)})=\frac{DJ^4K^2}{8(D+J^2)^4}\ .
\EEQ

We can then take a system in equilibrium at a temperature $T_e$, such that
 the configurations visited by the system at equilibrium
are the same as those out of equilibrium at temperature $T$.
This we call effective temperature.
In other words, fixing $t$, $T_e$ is defined as the temperature  at which 
the system at equilibrium would visit the same configurations visited by 
the system out of equilibrium  at temperature $T$,
with the same  frequency.


\subsubsection{The effective temperature employed in numerical approaches:
 the fragile case}

Following the approach found in literature \cite{KST,SKT,CR1}
for numerical simulations,
we can  define a  $T_{e1}^{(\rm is)}$ through the matching 
of the equilibrium and the out of equilibrium IS internal energy:
it is 
the one such that
\BEQ
E_{\rm eq}^{(\rm is)}(T_{e1}^{(\rm is)}(t))=E_{d}^{(\rm is)}(t)
\label{Te1def}
\EEQ

For our model 
it is possible to work out an analytic expression
 for such a  $T_{e1}^{(\rm is)}(t)$, at least   near the Kauzmann 
transition for the fragile glass case, linearizing in $T-T_0$. 

What we get is a parameter different from
the thermodynamic effective temperature (\ref{Te_T})
 that we got from three different approaches
(including the Fluctuation-Dissipation Ratio) in 
\cite{LN}.

For the fragile glass case we are not able to derive  any simple expression,
of the IS energy (\ref{Ed}), but we can in any case solve it exactly.
The results are shown in figures \ref{fig:Te_t_IS_40}-\ref{fig:Te_t_IS_41} for a
 given choice 
of the values of the interaction parameters of the model.
As one can see $T_{e1}^{(\rm is)}(t)$ comes out to be different from
$T_e(t)$ at any time decade.

As a matter of fact what we are comparing now with the average  $E_{d}(t)$
is a function $E_{eq}(T_{e1}^{(\rm is)})$  of the effective temperature alone,
while we know that out of equilibrium any proper thermodynamic
function cannot simply
depend on just one temperature as the thermodynamic function
of equilibrium systems do \cite{N00}. 
It is not surprising, thus, that the two functions do not coincide.


\subsubsection{The effective temperature employed in numerical approaches:
 the strong case}

For the strong glass case it is possible to work out a simple analytic expression
for the dynamical and the equilibrium IS energy. 
To do it we will expand near zero temperature up to second order in $T$.

 We underline that the thermodynamic effective temperature
given in (\ref{Te_T}) is also the expression of the 
effective temperature for the system
without constraint. What changes in that case  is the time
behavior of $m_2-m_1^2=\mu_2$, that is now given by (\ref{MU2S}),
 and its limit at equilibrium (see (\ref{SPeq2})).
In this case, where the analytic treatment is by far 
easier, we can give a short explicit expression for 
 $T_{e1}^{(\rm is)}(t)$:
\BEQ
T_{e1}^{(\rm is)}\simeq 
T+\frac{K D}{D+J^2}\delta\mu_2(t)
+\frac{J^4 K^2}{2(D+J^2)^3}T \delta\mu_2(t)+
\cO(T^3)+\cO\left(\delta\mu_2(t)^3\right)\ .
\label{Te1IS}
\EEQ
\noindent Here above terms of $\cO(T^2)$ and $\cO(\delta\mu_2(t)^2)$ cancel.
This $T_{e1}^{(\rm is)}(t,T)$ is obtained from (\ref{Te1def}) with
\BEA
\frac{ E_{\rm d}^{\rm is}(t)}{N}&&\simeq
-\frac{(H+J)^2}{2K}-L+\frac{T}{2}-\frac{J^4 K}{8 D(D+J^2)^2} T^2  
+\frac{K D}{2(D+J^2)}\delta\mu_2(t)
+\frac{J^4 K^2}{8 D(D+J^2)^2} T \delta\mu_2(t)
+\frac{D J^4 K^3}{8(D+J^2)^4} \delta\mu_2(t)^2
\nn
\\
&&\hspace*{5 cm}+\cO\left(T^3\right)
+\cO\left(T^2\delta\mu_2(t)\right)
+\cO\left(T\delta\mu_2(t)^2\right)
+\cO\left(\delta\mu_2(t)^3\right)
\label{EIS_S}
\EEA

If we expand  (\ref{Te_T}) in the same way we get:
\BEA
T_{e} &&= T+\Kt\delta\mu_2(t)=
T+ \frac{K D}{D+J^2} \delta\mu_2(t)+
\frac{T}{2}\left(\frac{J K}{D+J^2}\right)^2 \delta\mu_2(t)
+\frac{DJ^4K^3}{2(D+J^2)^4}\delta\mu_2(t)^2
\nn
\\
&&\hspace*{5 cm}+\cO\left(T^3\right)
+\cO(T^2\delta\mu_2(t))
+\cO\left(T\delta\mu_2(t)^2\right)
+ \cO\left(\delta\mu_2(t)^3\right)
\label{Te_exp}
\EEA

\begin{figure}[!htb]
\epsfxsize=10cm
\centerline{\epsffile{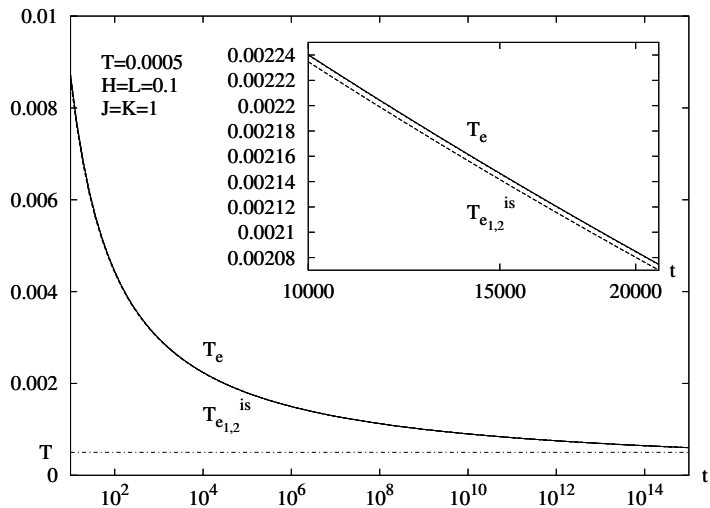}}
 \protect\caption{\small{Time evolution of the 
effective temperatures  at the heat bath temperature T = 0.0005
in the model with Arrhenius relaxation.
 The constants in the Hamiltonian (\ref{Hmodel}) are set to the following values: 
$K =  J  = 1$, $H = L = 0.1$.
The lower curve shows the effective temperature  (\ref{Te1IS}) got by matching
out of equilibrium and  equilibrium IS internal energy. 
To order $\delta\mu_2$ it coincides analytically
with the IS effective temperature (\ref{Te2IS}).  Second order 
differences are too small to appear in the plot.
The upper curve
is the behavior of (\ref{Te_T}), for systems at finite $T$.
}}
\label{fig:TeS_t_IS_01}
\end{figure}

\begin{figure}[!htb]
\epsfxsize=10cm
\centerline{\epsffile{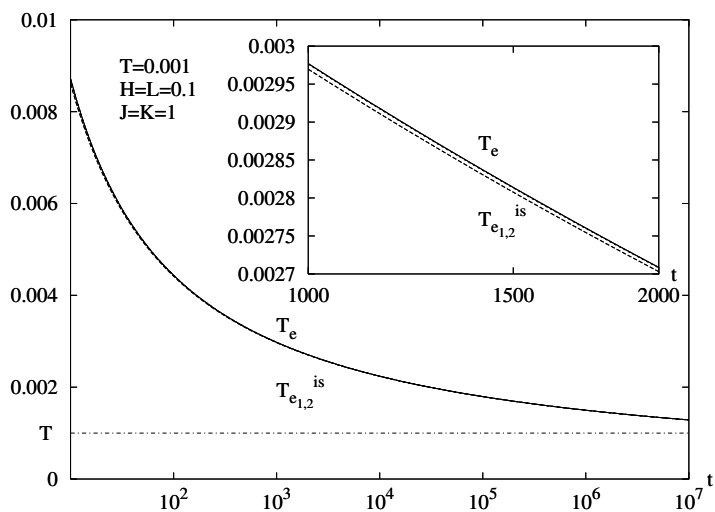}}
 \protect\caption{\small{The same effective temperatures, for the same 
choice of parameters as before are plotted for a different heat bath 
temperature: T = 0.001. Comparing the time scales of the two plots, we can
clearly observe the decreasing of the  
Arrhenius relaxation time to equilibrium $\tau_{\rm eq}$  that takes places 
rising the temperature.}}
\label{fig:TeS_t_IS_02}
\end{figure}

As we see from the formulae above and from figures 
\ref{fig:TeS_t_IS_01} and \ref{fig:TeS_t_IS_02},
 for a given choice
of the parameter values,
in the  case with Arrhenius relaxation 
$T_e$ and
$T_{e1}^{(\rm is)}$ are very similar. Their difference is
 one order of magnitude less than in the  
model with contrived dynamics.

\subsubsection{A more fundamental definition of the IS effective temperature}
Here we propose an alternative   way to identify an effective  temperature 
that maps the dynamics 
between inherent structures into a thermodynamic quantity. 
We  follow a quasi static approach using a partition sum,
 just as we did 
in the finite $T$ case. 
The aim is to be able to define an effective thermodynamic parameter
 for the IS dynamics and to compare it with
the $T_e$ given in (\ref{Te_T}).
Following exactly the same approach we used in \cite{LN} (see section
\ref{sec2:2T}), including the substitution of the real external field
$H$ with the effective one $H_{e2}^{({\rm  is})}$, we compute 
the partition function counting all the 
macroscopically equivalent 
IS's,
through which the system is evolving in this symbolic  dynamics, at a
given time $t$.

\begin{figure}[!htb]
\epsfxsize=10cm
\centerline{\epsffile{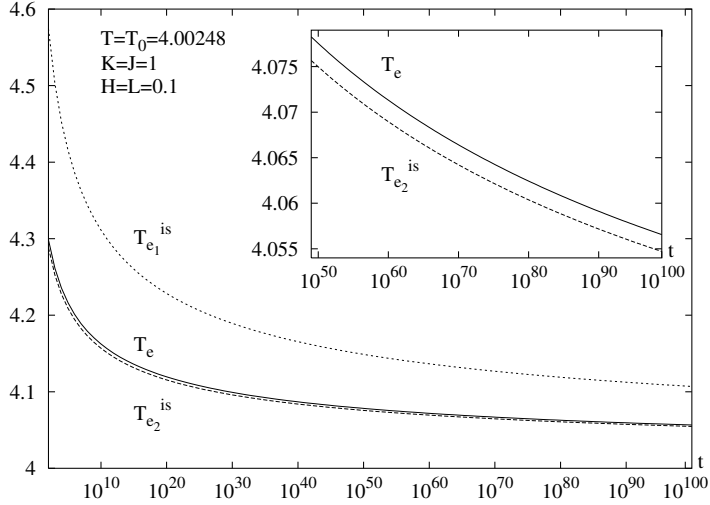}}
 \protect\caption{\small{Effective temperatures vs. t at the heat bath temperature $T = 4.00248$, equal to the Kauzmann temperature.
 The constants in the Hamiltonian (\ref{Hmodel}) are set to the following values: $K = J = 1$, $H = L = 0.1$. The constraint constant is $m_0 = 5$.
The upper curve shows the effective temperature got by matching
out of equilibrium and  equilibrium IS internal energy. The one in the middle
is the behavior of (\ref{Te_T}), for systems at finite $T$,
 and the lowest one is
the IS effective temperature (\ref{Teis_t}).}}
\label{fig:Te_t_IS_40}
\end{figure}

\begin{figure}[!htb]
\epsfxsize=10cm
\centerline{\epsffile{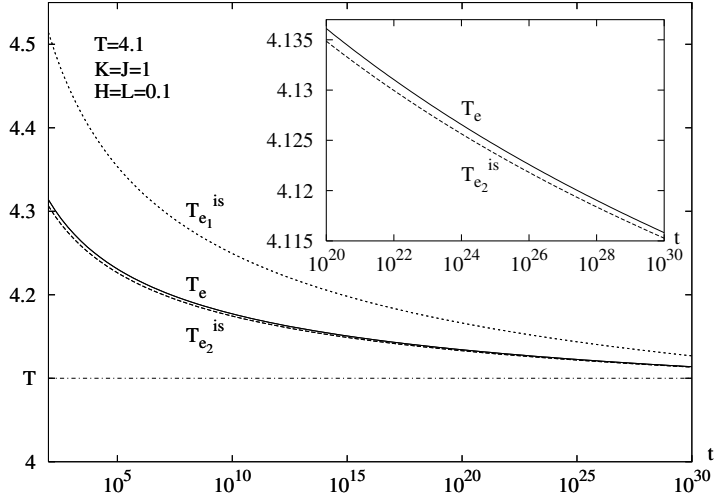}}
 \protect\caption{\small{
The same effective temperatures, for the same 
choice of parameters as before are plotted for a different heat bath 
temperature:$T = 4.1$.}}
\label{fig:Te_t_IS_41}
\end{figure}

\BEA
Z_e^{(\rm is)}(m_1,m_2)&&=\int{\cal{D}}x
\exp{\left[-\beta^{(\rm is)}_{e2} 
\H_{\rm is}\left(\{x_i\};T,H_{e2}^{(\rm is)}\right)\right]}
\delta\hspace*{-1mm}\left(N m_1-\sum_i x_i\right)
\delta\hspace*{-1mm}\left(N m_2-\sum_i x_i^2\right)=
\\
&&=
\exp{\left\{-\beta_{e2}^{(\rm is)} N
\left[\frac{K}{2}m_2-H_{e2}^{(\rm is)}m_1
-{\overline{w}}_{\rm is}-\frac{T_{e2}^{(\rm is)}}{2}
\log\left(m_2-m_1^2\right)\right]\right\}}=
\\
&&\simeq
\exp{\left\{-\beta_{e2}^{(\rm is)}
\left[{\cal{H}}_{\rm is}\left(m_1,m_2;T,H_{e2}^{(\rm is)}\right)
-T_{e2}^{(\rm is)}
{\cal{I}}\left(m_1,m_2\right)\right]\right\}} \ .
\EEA
\noindent $\beta^{(\rm is)}_{e2}=1/T_{e2}^{(\rm is)}$
 and $H_{e2}^{(\rm is)}$ are  parameters describing the behavior of
the  system going only through IS's.
Minimizing the free energy 
$F_e^{(\rm is)}\equiv -T_{e2}^{(\rm is)}\log Z_e^{(\rm is)}$ with respect to
$m_{1,2}$
we get:
\BEA
&&T_{e2}^{(\rm is)}=\Kt_{\rm is}\left(m_1,m_2\right)
\left[m_2-m_1^2\right] \ ,
\label{TeIS2}
\\
&&H_{e2}^{(\rm is)}=H-\Kt_{\rm is}\left(m_1,m_2\right)\mu_1 \ .
\EEA
By inserting the time dependent values of $m_1$ and $m_2$ 
we now look at the time evolution of the effective temperature (\ref{TeIS2})
for large times, in the aging regime, and we compare it with the 
behavior of the thermodynamic 
effective temperature  (\ref{Te_T}).

For the dynamically constrained model, 
for  $t\to\infty$, $T_{e2}^{(\rm is)}\to T$ (if $T > T_0$).
When $t_0\ll t <\infty$, however, the way the effective 
temperature approach the heat bath temperature
 is different from the behavior
 (\ref{Te_T})  of $T_e$,  found in the  case at finite temperature.
For a comparison
their first order expansions are:
\BEA
&&T_{e2}^{(\rm is)}\simeq T +  \left(1+T\frac{\Kt_{{\rm is},\tinf}(T)
Q^{(\rm is)}_{\tinf} J^2}
{2 D (1+Q^{(is)}_{\tinf} D)}\right) \Kt_{{\rm is},\tinf}(T) \ \delta \mu_2(t)
\label{Teis_t}
\\ 
&&T_e\simeq T +  \left(1+T\frac{\Kt_{\tinf}(T) Q_{\tinf} J^2}
{2 D (1+Q_{\tinf} D)}\right) \Kt_{\tinf}(T) \ \delta \mu_2(t)
\label{Te_t}
\EEA
\noindent
where
\BEA
&&
\Kt_{{\rm is},\tinf}(T)=\lim_{t\rightarrow \infty}
\Kt_{\rm is}\left(m_1(t),m_2(t);T\right) \ ,\\
&&
\Kt_{\tinf}(T)=\lim_{t\rightarrow \infty}
\Kt\left(m_1(t),m_2(t);T\right) \ ,
\nn
\\
&&
Q^{(\rm is)}_{\tinf}=\lim_{t\rightarrow \infty}\frac{J^2 D}
{\Kt_{\rm is}^3 w_{\rm is}^3} \ ,\\
&&
Q_{\tinf}=\lim_{t\rightarrow \infty} Q=
\lim_{t\rightarrow \infty} \frac{J^2 D}{\Kt^3 w\left( w+T/2\right)^2}\ .
\EEA
The time dependent variable 
$\delta\mu_2(t)$ is introduced in  (\ref{deltamu2F})  in both cases
(apart from the parameter $t_0$ influencing only the  short times) while
the coefficients in front of it are different at
 any temperature, including $T_0$.
In the fragile case, thus, this second IS effective temperature does not coincide 
with
$T_{e1}^{(\rm is)}$ and it  is
much nearer, at any time, to  (\ref{Te_t}).
However, even if this $T_{e2}^{(\rm is)}$ is conceptually  more properly
chosen than the one defined matching out of equilibrium 
energy at temperature $T$ and equilibrium energy at temperature $T_e^{(is)}$,
 we still do not get the same parameter describing 
the finite $T$ dynamics in a thermodynamic frame. 
The inherent structure approach gives thus a good approximation
but  is nevertheless never analytically correct.

To show how good this approximation is we can take 
as an instance a certain realization of the model
with given values of the ``fields'' and ``coupling constants''.
We plot in figures \ref{fig:Te_t_IS_40} and
\ref{fig:Te_t_IS_41} the behavior of $T_{e1}^{(\rm is)}(t)$, 
$T_{e2}^{(\rm is)}(t)$
and $T_e(t)$
at heat-bath temperatures equal to and just above the Kauzmann temperature.

For the strong glass case we also expand for
 temperatures near to zero and for long time and we get:
\BEA
T_{e2}^{(\rm is)}(t) &&= T+\Kt_{\rm is}\delta\mu_2(t)=
 T+ \frac{K D}{D+J^2} \delta\mu_2(t)+
\frac{T}{2}\frac{J^4 K^2}{\left(D+J^2\right)^3} \delta\mu_2(t)
+\frac{D J^4 K^3}{2(D+J^2)^4}\delta\mu_2(t)^2
\\
\nn
&&\hspace*{5 cm}
+\cO(T^3)+\cO(T^2\delta\mu_2(t))
+\cO(T\delta\mu_2(t)^2)+\cO\left(\delta\mu_2(t)^3\right)
\label{Te2IS}
\\
&& = T_e(t) -\frac{DJ^2K^2}{2(D+J^2)^3}T \delta\mu_2(t) 
+\cO\left(\delta\mu_2(t)^3\right)
+\cO(T^3)+\cO(T^2\delta\mu_2(t))
+\cO(T\delta\mu_2(t)^2)
\\
&& = T_{e1}^{(\rm is)}(t) +\frac{DJ^4K^3}{2(D+J^2)^4}\delta\mu_2(t)^2 +
\cO(T^3)+\cO(T^2\delta\mu_2(t))
+\cO(T\delta\mu_2(t)^2)+\cO\left(\delta\mu_2(t)^3\right)
\EEA
\noindent where $\delta\mu_2$ is given
by (\ref{deltamu2S}).

The effective temperature $T_e$ mapping the dynamics of the system evolving
at finite  temperature $T$ have the same behavior of $T_{e2}$
in approaching the
heat-bath temperature up to order  $T\delta\mu_2(t)$
where they start deviating one from the other.
For a quenching to zero temperature the two effective temperatures
coincide. 
Moreover,  due to the simplicity of the model, the IS 
effective temperature $T_{e2}^{(\rm is)}$ is equal to
with $T_{e1}^{(\rm is)}$ given in (\ref{Te1IS}) up to order $
\delta\mu_2(t)$ in time and up to order $T^3$ in 
temperature.

 \renewcommand{\thesection}{\arabic{section}}
\section{Conclusions}
\setcounter{equation}{0}\setcounter{figure}{0}
\renewcommand{\thesection}{\arabic{section}.}

In this paper we consider a model
 that owns all the basic properties 
of a glass,  built 
 by processes evolving on two  well separated time scales,
 representing the $\alpha$ and $\beta$ processes taking place in real
glassy materials \cite{LN}.
The decoupling of time scales is fundamental  for 
a generalization of equilibrium thermodynamics to systems 
far from equilibrium.

We take into account two different versions of the model given by 
(\ref{Hmodel}). One leading to the description of 
a fragile glass having a non-zero Kauzmann temperature and the 
other one representing a strong glass.

Using
 a particular  Monte Carlo dynamics
 and  developing it 
analytically, thus having the opportunity of probing it in more detail with
respect to a numeric study,
 we found equations of motion that are in all respect 
 those typical of glass relaxation.

In the strong glass case we apply exactly the same 
parallel Monte Carlo dynamics used in 
\cite{BPPR,BPR,NPRL98,N00}, finding an Arrhenius relation between
 the relaxation time of the slow processes $\{x_i\}$ 
and the temperature.

In the fragile glass  case
the model is provided with a constraint
applied to the harmonic oscillator dynamics, i. e. to the slow 
processes dynamics, in order to  reproduce the behavior of a 
good fragile glass former. In \cite{LN},
by means of a Monte Carlo constrained dynamics, we identified
the Kauzmann temperature with 
 the one, $T_0$, at which  the constraint is reached, asymptotically, 
for the first time in a cooling experiment from high temperature.
There we showed how
the  thermodynamic phase transition \cite{KAUZ} is characterized, 
that takes place
 due to the breaking of the ergodicity in the  landscape of our model,
rich of degenerate minima.

In this work we  carried out the inherent structure approach.
In both dynamical  models,
 decreasing the temperature, the  free energy local minima 
do not  split into smaller
 local minima, just like in 
the $p$-spin model in zero magnetic field \cite{CS},
because every allowed  configuration of harmonic oscillators
is and stays an inherent structure at any
temperature.
Consequently   we can set a one-to-one 
correspondence  between the minima of the
free energy and the ones of the potential energy
 (i. e. the inherent structures).
Because of this exact correspondence the dynamics through inherent structures 
should be a valid symbolic dynamics for the real system,
i.e. at a finite heat-bath temperature $T$.
At least, it would significantly represents
 the actual dynamics 
if the inherent structures  are visited with
the same frequency with which the corresponding free energy 
minima at finite $T$ are visited.

We   defined  the steepest descent procedure for the model, that
is the minimization of the effective Hamiltonian  (\ref{Heff})
appearing in the partition function with respect to the spins,
 i. e. the fast relaxing variables.
Due to the minimization any explicit dependence on $T$ disappears.
The configurational entropy for inherent structures was computed from the 
logarithm of the Jacobian of the transformation of variables
${\cal{D}}x \to dm_1dm_2$, 
and thus it was the same of the 
exact finite $T$ approach. 
In our models, then,  any  configuration of harmonic oscillators
$\{x_i\}$ (for the fragile glass model every configuration
allowed by the constraint),
 is also an inherent structure.
Although the models we considered are conceptually very simple and 
without interactions, as compared to another approach proposed for
systems with interacting 
discrete spins where the IS scheme breaks down \cite{BMEPL00},
our setup seems to be more physical since it is intimately based on time
scale separation between fast and slow processes.
A direct consequence of this time scale separation is that we encounter
a both mathematically and physically well defined 
configurational entropy, whereas this observable suffers from principle
difficulties in \cite{BMEPL00}.

We can  take a system in equilibrium at an effective
 temperature $T_e^{\rm (is)}$, such that
 the configurations visited by the system at equilibrium
are the same as those out of equilibrium at temperature $T$.
First we  defined an effective temperature through the matching 
of the equilibrium and the out of equilibrium internal energy
of the inherent structures
(the one such that $E(T_e(t))=E_{d}(t)$).
For the strong glass model this effective temperature
almost coincides with the finite temperature $T_e$ provided
that  the temperature at which  the system is 
quenched is not too high (as far as terms of $\cO(T\delta\mu_2(t))$
are negligible they are equal).
On the contrary, when the constraint is set and the contrived 
Monte Carlo dynamics is applied, we found that the thus derived 
effective temperature $T_{e1}^{(\rm is)}$ 
 is quite different from the effective temperature
that we were able to identify in the finite $T$ dynamics.
Therefore
 we proposed a new  definition
  following a quasi static approach. 
In this way we computed 
the partition function counting all the 
macroscopically equivalent 
inherent structures,
through which the system is evolving in this symbolic  dynamics, at a
given time $t$.
Even though the result we get is much more similar to the finite $T$
dynamics effective temperature (numerically speaking the difference is 
one order of magnitude smaller), yet it is analytically
different, indicating that the inherent structure scheme can
 only be an approximation to what happens
in the realistic dynamics of the system.
As a consequence, 
also the derivation of out of equilibrium 
thermodynamic quantities (e.g. the configurational entropy) 
obtained making use of this approach could suffer of a systematic deviation
from the exact result.

 \renewcommand{\thesection}{\arabic{section}}
\section*{Acknowledgments}
\setcounter{equation}{0}\setcounter{figure}{0}
\renewcommand{\thesection}{\arabic{section}.}
We thank F. Ritort for suggestions and a careful reading of the manuscript.
We also thank A. Crisanti and F. Sciortino for useful discussions.
The research of L. Leuzzi is supported by FOM (The Netherlands).

\renewcommand{\thesection}{\arabic{section}}
 \appendix{}
\section*{Strong glass dynamics}
\setcounter{equation}{0}\setcounter{figure}{0}
\renewcommand{\thesection}{\arabic{section}.}
\renewcommand{\theequation}{A.\arabic{equation}}
\label{app:A}

In this appendix we present the Monte Carlo dynamics of the observables
$\mu_1$ and $\mu_2$, functions of the slow relaxing harmonic 
oscillators $\{x_i\}$ through
\BEQ 
m_1=\frac{1}{N}\sum_i x_i \hspace*{2cm} , \hspace*{2cm}
 m_2=\frac{1}{N}\sum_i x_i^2 \ ,
\EEQ
in the case where the model (\ref{Hmodel}) is not subjected to
any constraint on its $\{x_i\}$ configurations.

Let us recall the definitions, given in section \ref{dyn},
\BEA
&&\mu_1\equiv \frac{\Ht}{\Kt}-m_1
\\
&&\mu_2\equiv m_2-m_1^2
\EEA

In this notation the average and the variance (\ref{def:MCave})
of the gaussian distribution 
$p(x|m_1,m_2)$ (see (\ref{PROBDIST}))
of the possible changes in energy during the Monte Carlo
dynamics become:
\BEQ
\overline{x}=\frac{\Delta^2\Kt}{2} \hspace*{1cm} \ , \hspace*{1 cm}
\Delta_x=\Delta^2\Kt^2\left(\mu_2+\mu_1^2\right) \ .
\EEQ
We remember here that $x$ is the difference (\ref{x}) between the energy of the configuration proposed  for the exchange
and the energy of the actual configuration.
$\Delta$ is fixed. 

To shorten the following  expressions 
we also define the parameter
\BEQ
\alpha\equiv\frac{\overline{x}}{\sqrt{2\Delta_x}}=
\sqrt{\frac{\Delta^2}{8\left(\mu_2+\mu_1^2\right)}}
\label{alpha}
\EEQ

The two  basic quantities that have to be computed in
order to solve the dynamic equations for $\mu_1$ and $\mu_2$ are
 the acceptance rate of the Monte Carlo
updating:
\BEQ
A(t)\equiv\int d x \ W(\beta x)\  p(x|m_1,m_2)
\EEQ
and the rate of change of the energy of the system:
\BEQ
I_1(t)\equiv\int  d x\   x \ W(\beta x)\  p(x|m_1,m_2)
\EEQ

Defining the auxiliary function:
\BEQ
f(t)\equiv \overline{x}\beta\exp\left(-\beta\overline{x}+\frac{\beta^2\Delta_x}{2}\right)\mbox{erfc}\left(
\sqrt{\frac{\Delta_x}{\overline{x}}}\beta
-\alpha\right) \ ,
\EEQ
\noindent
where
\BEQ
\mbox{erfc}(a)\equiv\frac{2}{\sqrt{\pi}}\int_a^{\infty} dz\ e^{-z^2} \ ,
\EEQ
we can write down the exact expressions for $A$ and $I_1$ as
\BEA
&& A=\frac{1}{2}\left[\mbox{erfc}(\alpha)+\frac{f}{\beta\overline{x}}\right]
\\
&&I_1	=\frac{\overline{x}}{2}\left[\mbox{erfc}(\alpha)+\left(1-
\frac{\beta \overline{x}}{2\alpha^2}\right)\frac{f}{\beta{\overline{x}}}\right] \ .
\EEA

The Monte Carlo equations of motion for $\mu_1$ and $\mu_2$ are 
formally the same found for the fragile glass case \cite{LN}:
\BEA
&&\dot{\mu_1}=-J Q\int  d x \ x\ W(\beta x)\ p(x|m_1,m_2)
-\left(1+DQ\right)\int  d x \ {\overline{y}}(x)\ W(\beta x)\ p(x|m_1,m_2)
\\
&&\dot{\mu_2}=\frac{2}{\Kt}\int  d x\  x\ W(\beta x)\ p(x|m_1,m_2)
+2\mu_1\int  d x \ {\overline{y}}(x)\ W(\beta x)\ p(x|m_1,m_2)
\EEA
From (\ref{def:MCave}) we know $\overline{y}(x)$ and 
 we can rewrite it as a function of the above defined $\alpha$:
\BEQ
{\overline{y}(x)}=4\alpha^2\mu_1\left(1-\frac{x}{\overline{x}}\right)
\EEQ
Using this we get
\BEA
\dot{\mu_1}&=&-\left(J Q-\frac{8 \alpha^2\mu_1(1+DQ)}{\Delta^2\Kt}\right)I_1(t)
-4\alpha^2\mu_1(1+DQ)A(t)
\label{eqMU1}
\\
\dot{\mu_2}&=&\frac{2}{\Kt}\left(1-\frac{8\alpha^2\mu_1^2}{\Delta^2\Kt}
\right)I_1(t)+8\alpha^2\mu_1^2A(t)
\label{eqMU2}
\EEA


\subsection{ Dynamics in the aging regime: zero temperature}

First of all we solve the equation of motion for $\mu_2$  at $T=0$,
 neglecting  terms of order $\mu_1^2$ with respect to those of order $\mu_2$.
For long times is $\alpha\gg 1$. We can, then, 
 expand $I_1(t)$ for large $\alpha$,
getting
\BEQ
I_1(t)\simeq-\frac{e^{-\alpha^2}}{2\alpha\sqrt{\pi}}\frac{\Delta^2\Kt}
{4 \alpha^2}
\label{I10}
\EEQ

Equation (\ref{eqMU2}) becomes then
\BEQ
\dot\mu_2\simeq -\frac{e^{-\alpha^2}}{2\alpha \sqrt{\pi}}\frac{\Delta^2}{4 \alpha^2}
\label{eqmu2} \ ,
\EEQ
\noindent otherwise written as 
\BEQ
\dot\alpha\simeq\frac{e^{-{\alpha^2}}}{\sqrt{\pi}}
\label{alphaEQ}
\EEQ
\noindent or
\BEQ
\dot\mu_2\simeq-2\mu_2^{3/2}\frac{\exp\left(-\frac{\Delta^2}{8\mu_2}\right)}{\sqrt{\pi}}
\EEQ

At $T=0$, the solution in the aging regime, expressed in
$\mu_2\simeq\Delta^2/(8\alpha^2)$, turns out to be:
\BEQ
\mu_2(t)\simeq \frac{\Delta^2}{8}\frac{1}{\log{\frac{2 t}{\sqrt{\pi}}}+
\frac{1}{2}\log\log{\frac{2 t}{\sqrt{\pi}}}} \ .
\label{MU2Ssol0}
\EEQ

Always at zero temperature the leading order of the expansion of the 
acceptance rate $A$ is , for $\alpha \gg 1$, 
\BEQ
A\simeq \frac{e^{-\alpha^2}}{2\alpha\sqrt{\pi}}
\EEQ
Combining this with (\ref{I10}), the Monte Carlo equation  of motion (\ref{eqMU1})
takes the form
\BEQ
\dot\mu_1=\frac{e^{-\alpha^2}}{2\alpha\sqrt{\pi}}\left\{\frac{JQ\Delta^2\Kt}{4\alpha^2}-
2\mu_1(1+DQ)\left(2\alpha^2+1\right)\right\}
\label{eqmu1}
\EEQ

Dividing (\ref{eqmu1}) for (\ref{eqmu2}) we can write down a differential
equation for $\mu_1$ as a function of $\mu_2$:
\BEQ
\frac{d \mu_1}{d \mu2}\simeq
16(1+DQ)\frac{\alpha^4}{\Delta^2}\mu_1-JQ\Kt
\label{mu1mu20}
\EEQ
\noindent where we have neglected terms of order $1/\alpha^2$ with
respect to those of order $1$.
In the adiabatic approximation, obtained by neglecting the left hand side,
the solution of (\ref{mu1mu20}) turns out to be
\BEQ
\mu_1\simeq \frac{4 J Q\Kt}{\Delta^2 (1+DQ)} \mu_2^2
\label{mu1Ssol0}
\EEQ
At zero temperature and for long times, one thus have
 $\mu_1\sim \mu_2^2\ll \mu_2$.


\subsection{Dynamics in the aging regime: $T>0$} 

If $T$ is above zero, the leading order of the expansion of $A$ and $I_1$
 for large times ($\alpha\gg 1$) and small temperature are:
\BEA
&&A\simeq\frac{e^{-\alpha^2}}{2\alpha\sqrt{\pi}}\frac{1}{1-\frac{4T\alpha^2}{\Delta^2\Kt}}
\label{A_T}
\\
&&I_1\simeq \frac{e^{-\alpha^2}}{2\alpha\sqrt{\pi}}
\frac{\Delta^2 \Kt}{4\alpha^2}
\frac{1-\frac{8T\alpha^2}{\Delta^2\Kt}}
{\left(1-\frac{4T\alpha^2}{\Delta^2\Kt}\right)}
\label{I1_T}
\EEA
\noindent where the terms $T\alpha^2$ are of $\cO(1)$. Indeed it is 
\BEQ
\frac{8T\alpha^2}{\Delta^2\Kt}=\frac{8T}{\Delta^2\Kt}\frac{\Delta^2}{8\mu_2}
=\frac{\bamu}{\mu_2}=\frac{1}{1+\frac{\delta\mu_2}{\bamu}} \ ,
\EEQ
(see the definition of $\alpha$ (\ref{alpha})) so that
\BEQ
\lim_{t\to\infty}\frac{8T\alpha^2}{\Delta^2\Kt}=1\ , \hspace*{ 1cm}  \ T>0 \ .
\EEQ
In this notation (\ref{A_T}) and (\ref{I1_T}) can be rewritten also as
\BEA
&&A=\frac{e^{-\alpha^2}}{\alpha\sqrt{\pi}}
\frac{1+\frac{\delta\mu_2}{\bamu}}
{1+2\frac{\delta\mu_2}{\bamu}}
\\
&&I_1=-\frac{e^{-\alpha^2}}{\alpha\sqrt{\pi}}\frac{\Delta^2\Kt}{\alpha^2}
\frac{\delta\mu_2}{\bamu}\frac{1+\frac{\delta\mu_2}{\bamu}}
{\left(1+2\frac{\delta\mu_2}{\bamu}\right)^2}
\EEA
and the Monte Carlo equations are now
\BEA
&&\dot\mu_1\simeq\frac{e^{-\alpha^2}}{\alpha\sqrt{\pi}}
\frac{1+\frac{\delta\mu_2}{\bamu}}
{1+2\frac{\delta\mu_2}{\bamu}}\left[-4\alpha^2(1+DQ)\mu_1
+\frac{JQ\Delta^2\Kt}{2\alpha^2}\frac{\delta\mu_2}{\bamu}
\frac{1}{1+2\frac{\delta\mu_2}{\bamu}}\right]
\label{eqmu1T}
\\
&&\dot\mu_2\simeq\frac{e^{-\alpha^2}}{\alpha\sqrt{\pi}}
\frac{\Delta^2}{\alpha^2}\frac{\delta\mu_2}{\bamu}
\frac{1+\frac{\delta\mu_2}{\bamu}}
{\left(1+2\frac{\delta\mu_2}{\bamu}\right)^2}
\label{eqmu2T}
\EEA
The solution to (\ref{eqmu2T}) is, to leading order,
\BEQ
\mu_2(t)\simeq\frac{\Delta^2}{8}\frac{1}{\log{\frac{2 t}{\sqrt{\pi}}}} \ .
\EEQ
The behaviour of $\mu_1$ comes out to be
\BEQ
\mu_1\simeq\frac{4JQ\Kt}{\Delta^2(1+DQ)}\mu_2^2 \ \frac{\delta\mu_2}{\bamu}
\frac{2}{1+2\frac{\delta\mu_2}{\bamu}} \ .
\label{mu1SsolT}
\EEQ
\noindent Notice that this vanishes when equilibrium  is approached, since
then $\delta\mu_2 \to\ 0$. 

For times even longer than the time scale of the aging regime, 
the system finally relax, exponentially fast, to equilibrium.
The equilibrium value of $\mu_2$ is known from the statics (see section \ref{sec2}),
as
\BEQ
\bamu=\frac{T}{\Kt_{\tinf}(T)}
\EEQ
\noindent where the explicit expansion of $\Kt_{\tinf}(T)$ in temperature is shown in
(\ref{KtSexp}).
The asymptotic value of $\alpha$ is, from its definition (\ref{alpha}) and
taking the first order expansion in $T$,
\BEQ
\alpha(T)=\sqrt{\frac{\Delta^2}{8\bamu(T)}}\simeq\sqrt{\frac{A_{\rm s}}{T}}
\label{alphaI}
\EEQ
with
\BEQ
A_{\rm s}\equiv \frac{\Delta^2 K D}{8(D+J^2)} \ .
\EEQ

From the equations of motion studied above (look for instance at
(\ref{alphaEQ}))  we find for the relaxation time to 
equilibrium
\BEQ
\tau_{\rm eq}\propto e^{\alpha^2}\ .
\EEQ
Using (\ref{alphaI}) this is nothing else than the Arrhenius law 
\BEQ
\tau_{\rm eq}\sim \exp\left(\frac{A_{\rm s}}{T}\right)\ .
\EEQ

\addcontentsline{toc}{chapter}{}

\end{document}